\documentclass[aps,prb,twocolumn,showpacs]{revtex4}
\usepackage{stmaryrd}
\usepackage{amssymb}

\usepackage{pifont}
\usepackage{txfonts}
\usepackage{graphicx}

\begin{document}

\title{Superconductivity at 36 K in Gadolinium-arsenide Oxides GdO$_{1-x}$F$_{x}$FeAs}

\author{Peng Cheng, Lei Fang, Huan Yang, Xiyu Zhu, Gang Mu, Huiqian Luo, Zhaosheng Wang and Hai-Hu Wen}\email{hhwen@aphy.iphy.ac.cn }

\affiliation{National Laboratory for Superconductivity, Institute of
Physics and Beijing National Laboratory for Condensed Matter
Physics, Chinese Academy of Sciences, P.O. Box 603, Beijing 100190,
People's Republic of China}

\begin{abstract}
In this paper we report the fabrication and superconducting
properties of GdO$_{1-x}$F$_{x}$FeAs. It is found that when x is
equal to 0.17, GdO$_{0.83}$F$_{0.17}$FeAs is a superconductor with
the onset transition temperature T$_{c}^{on}\approx$ 36.6K.
Resistivity anomaly near 130K was observed for all samples up to x =
0.17, such a phenomenon is similar to that of
LaO$_{1-x}$F$_{x}$FeAs. Hall coefficient indicates that
GdO$_{0.83}$F$_{0.17}$FeAs is conducted by electron-like charge
carriers.

\end{abstract}
 \maketitle

\section{INTRODUCTION}

Since the discovery of superconductivity in iron-based layered
quaternary compound LaOFeP\cite{10K}, extensive efforts have been
devoted to find new superconductors among this
system.\cite{10K,LOMP2,26K,Pr,Nd,Ce,Sr,XHChen,NLWang,LeiFang} It is
found that with the replacement of $P$ by $As$ and partial
substitution of O with F, LaO$_{1-x}$F$_{x}$FeAs changes into
superconducting state below $T_c \approx 26$.\cite{26K} Subsequently
superconductivity at 25 K was also observed in
La$_{1-x}$Sr$_{x}$OFeAs in which no F was added into the sample,
therefore it was a hole-doped superconductor.\cite{Sr} More
recently, superconductors LnO$_{1-x}$F$_{x}$FeAs with light
rare-earth substitution (Ln=Ce, Pr, Sm) were realized and
superconducting transition temperature(T$_{c}$) was raised to
52K.\cite{XHChen,Ce,Pr,Nd} As to the heavy rare-earth element,
however, single phase could not be easily formed and no
superconducting state was observed below 2K. As to the element
Gadolinium which locates near the heavy rare-earth element,
experimentally a drop of resistivity was observed below 10 K but
with a residual resistivity down to 2 K.\cite{NLWang} So it is worth
exploring further that whether GdO$_{1-x}$F$_{x}$FeAs is also a
superconductor with much higher $T_c$. In this study, we report the
superconducting properties of GdO$_{0.83}$F$_{0.17}$FeAs, the onset
transition temperature T$_{c}^{on}$ is about 36.6 K.

\section{EXPERIMENT}

Polycrystalline samples GdO$_{1-x}$F$_{x}$FeAs (x=0.12, 0.15, 0.17)
were synthesized by conventional solid state sintering method. The
raw materials are all with high purity(Gd$_{2}O_{3}$ 99.99\%,
GdF$_{3}$ 99.99\%, Fe 99.95\%, As 99.99\%, Gd 99.99\%). The detailed
synthesis method is the same as that in the papers we reported
previously.\cite{XiyuZhu,LeiFang}. The as-sintered pullet is
concrete ceramic-like with dark-brown surface. X-Ray diffraction
measurement was performed at room temperature using an
MXP18A-HF-type diffractometer with Cu-K$_{\alpha}$ radiation from
10$^\circ$ to 80$^\circ$ with a step of 0.01$^\circ$. The
magnetization measurements were carried out on a Quantum Design
superconducting quantum interference device (SQUID) magnetometer.
The electrical resistivity and Hall coefficient were measured by a
Physical Property Measurement System (PPMS, Quantum Design) with a
standard six-probe method.

\begin{figure}
\includegraphics[width=9cm]{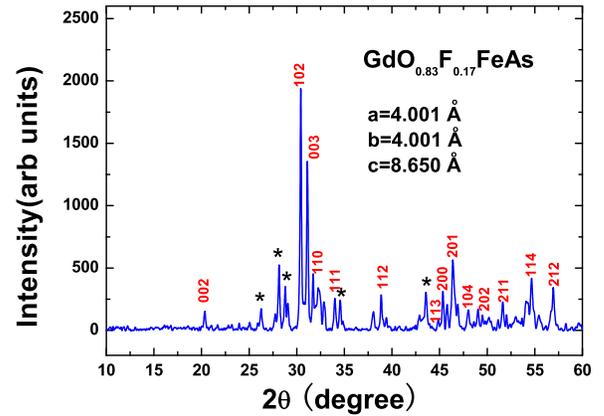}
\caption {(color online) XRD pattern of GdO$_{1-x}$F$_{x}$FeAs
(x=0.17). It is clear that the dominant phase is
GdO$_{1-x}$F$_{x}$FeAs. The asterisks mark the peaks from the
impurity phase.} \label{fig1}
\end{figure}

Fig.1 shows the X-ray diffraction (XRD) pattern of the sample
GdO$_{0.83}$F$_{0.17}$FeAs. The pattern can be indexed in tetragonal
space group with $a$ = $b$ = 4.001 {\AA} and $c$ = 8.650 {\AA}.
Obviously the phase is dominated by GdO$_{1-x}$F$_{x}$FeAs, though
minor impurity phases still exist as marked by the asterisks.  Such
impurity phases could be caused by the inadequate sintering
temperature 1160 $^{\circ}$ C in our experiment. The indices of the
crystal lattice we obtained are consistent with the counterparts of
LnO$_{1-x}$F$_{x}$FeAs (Ln=Ce, Pr, Sm)\cite{XHChen,Ce,Pr,Nd}.

Fig.2 shows the temperature dependence of DC magnetization  for
sample GdO$_{0.83}$F$_{0.17}$FeAs. The diamagnetic signal appears
below 22 K, a simple estimation on the magnetization at 2K reveals
that the superconducting volume fraction is more than 40\%. It
should be noted that the response to magnetic field in the normal
state of GdO$_{0.83}$F$_{0.17}$FeAs is paramagnetic with a minor
value compared to diamagnetic value and such a contribution to
magnetization is subtracted as a background (the same to ZFC and FC
curve).

\begin{figure}
\includegraphics[width=9cm]{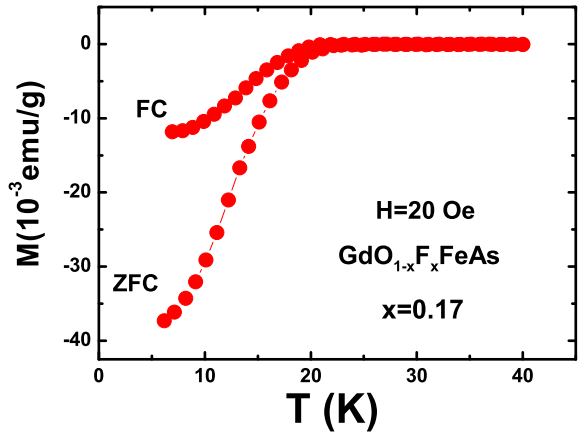}
\caption {(color online) DC magnetization of
GdO$_{0.83}$F$_{0.17}$FeAs measured in the zero-field-cooled
       (ZFC) and field-cooled (FC) processes. A diamagnetic signal is easily observed at about 22 K } \label{fig4}
\end{figure}

 The temperature dependence of resistivity of all
three samples are shown in Fig.3. We can see that the resistivity
shows an anomaly around 130 K, and this anomaly weakens with more F
doping, such an anomaly and corresponding evolution with F doping
have been observed in other Fe-based Arsenic compounds, however, the
anomaly did not happen in Nickel based Arsenic system. The obvious
disparity between Fe-based and Ni-based systems deserves to be
further studied. Both x = 0.15 and x=0.17 samples exhibit
superconducting transitions and zero-resistance at a lower
temperature. From the inset of Fig.3 we can see the onset drop of
resistivity about 36.6 K for x = 0.17 sample. As to
GdO$_{0.83}$F$_{0.17}$FeAs, a slight hump near 130 K is also
observed in resistivity curve, it suggests that T$_{c}$ could be
increased further as long as more Fluorine were doped into
GdO$_{1-x}$F$_{x}$FeAs. The transition width can also be narrowed in
a refined fabrication process in the future.

\begin{figure}
\includegraphics[width=9cm]{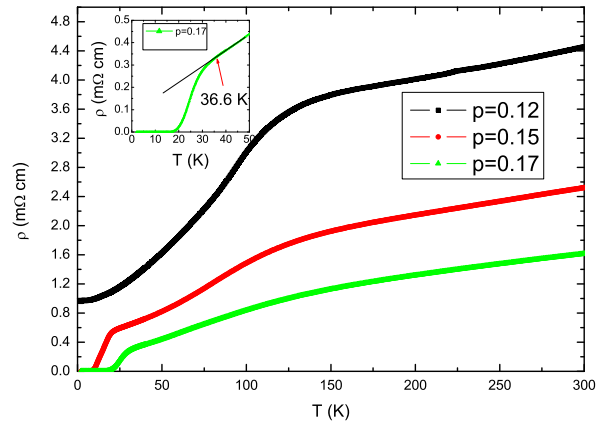}
\caption {(color online) The electrical resistivity vs temperature
for GdO$_{1-x}$F$_{x}$FeAs (x=0.12,0.15,0.17). The inset shows the
enlarged view of superconducting transition area for the sample x =
0.17. The onset transition temperature is defined at the point where
the resistivity starts to deviate from the normal state background
as marked by the straight line here.} \label{fig4}
\end{figure}

 Hall effect measurement for sample
GdO$_{0.83}$F$_{0.17}$FeAs was shown in Fig.4. The transverse
resistivities $\rho_{xy}$ above T$_{c}$ are all negative, indicating
that the normal state conduction of GdO$_{0.83}$F$_{0.17}$FeAs is
dominated by the electron-like charge carriers. The Hall coefficient
$R_H=\rho_{xy}/H$  changes slightly at high temperatures but drops
below 100 K. The value of $R_H$ is about -1$\times10^{-8}m^{3}/C$ at
100 K, compared with that of LaO$_{0.9}$F$_{0.1}$FeAs, the value of
Hall coefficient is similar.\cite{XiyuZhu} An estimation based on
the single band model gives a charge carrier density of 1$\times
10^{21}/cm^3$.

\begin{figure}
\includegraphics[width=9cm]{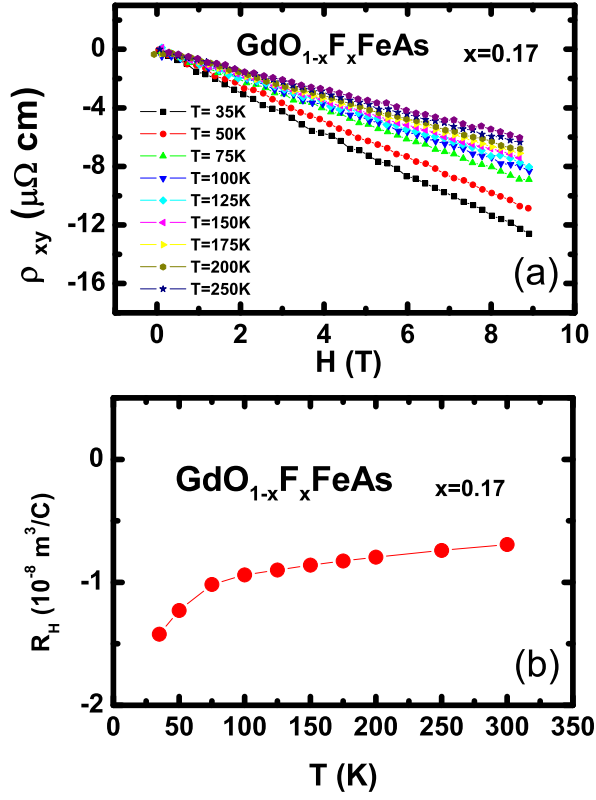}
\caption {(color online) (a) Transverse resistivity with relation of
magnetic field at different temperatures for
GdO$_{0.83}$F$_{0.17}$FeAs ; (b) Temperature dependence of Hall
coefficient for GdO$_{0.83}$F$_{0.17}$FeAs, the negative value
indicates that the charge carrier is electron type.} \label{fig4}
\end{figure}

\section{Concluding remarks}
In this study we report the fabrication and the superconducting
properties of GdO$_{1-x}$F$_{x}$FeAs, as x is equal to 0.17,
GdO$_{0.83}$F$_{0.17}$FeAs is a superconductor with the onset
transition temperature of about 36.6 K. Resistivity anomaly near 130
K was observed for all samples, which is similar to that of
LaO$_{1-x}$F$_{x}$FeAs. Hall coefficient suggests that
GdO$_{0.83}$F$_{0.17}$FeAs is conducted by electron-like charge
carriers.

\begin{acknowledgments}

This work is supported by the National Science Foundation of China,
the Ministry of Science and Technology of China (973 project No:
2006CB601000, 2006CB921802), and Chinese Academy of Sciences
(Project ITSNEM).
\end{acknowledgments}

\end{document}